\documentclass[runningheads]{llncs}

\usepackage{graphicx}
\usepackage[inline]{enumitem} 
\usepackage{booktabs}
\usepackage{adjustbox}
\usepackage[hidelinks]{hyperref} 
\usepackage[T1]{fontenc}
\usepackage[flushleft]{threeparttable}
\usepackage{refcount}
\usepackage{cite}

\usepackage{amssymb}
\usepackage{pifont}
\usepackage{array}

\newcolumntype{P}[1]{>{\centering\arraybackslash}p{#1}}

\usepackage[T1]{fontenc}
\usepackage[scaled=0.85]{beramono}
\usepackage{listings}
\begin{document}

\title{SmartReviews: Towards Human- and Machine-actionable Representation of\\ Review Articles}

\titlerunning{SmartReviews: Human- and Machine-actionable Review Articles}

\author{Allard Oelen\inst{1,2}\orcidID{0000-0001-9924-9153} \and
Markus Stocker\inst{2}\orcidID{0000-0001-5492-3212} \and
S\"oren Auer\inst{1,2}\orcidID{0000-0002-0698-2864}}

\authorrunning{Oelen et al.}

\institute{L3S Research Center, Leibniz University of Hannover, Hannover, Germany
\email{oelen@l3s.de}\and
TIB Leibniz Information Centre for Science and Technology, Hannover, Germany\\
\email{\{markus.stocker,soeren.auer\}@tib.eu}}

\newcounter{rtaskno}
\DeclareRobustCommand{\rtask}[1]{%
   \refstepcounter{rtaskno}%
   \textbf{R\thertaskno\label{#1}}}

\maketitle 

\begin{abstract}
Review articles are a means to structure state-of-the-art literature and to organize the growing number of scholarly publications. However, review articles are suffering from numerous limitations, weakening the impact the articles could potentially have. A key limitation is the inability of machines to access and process knowledge presented within review articles. In this work, we present SmartReviews, a review authoring and publishing tool, specifically addressing the limitations of review articles. The tool enables community-based authoring of living articles, leveraging a scholarly knowledge graph to provide machine-actionable knowledge. We evaluate the approach and tool by means of a SmartReview use case. The results indicate that the evaluated article is successfully addressing the weaknesses of the current review practices.

\end{abstract}

\section{Introduction}
\label{section:introduction}
As more scholarly articles are published every year~\cite{Jinha2010a}, methods and tools to organize published articles are becoming increasingly important~\cite{Klampfl2014}. Traditionally, review (or survey) articles are used to organize information for a particular research domain~\cite{Wee2016}. Research articles, also referred to as primary sources, present original research contributions. Review articles, or secondary sources, organize the research presented in the primary sources~\cite{Randolph2009a}. The importance of review articles becomes apparent in the fact that these articles are often highly cited~\cite{Wolmark2001} which indicates that they are valuable for the community. Although reviews are important, they suffer from several major weaknesses, which affect the potential impact review articles can have. For example, once review articles are published, they are generally not updated when new research articles become available. This results in reviews that are outdated soon after publication. Furthermore, scholarly articles are not machine-actionable, which prevents machines from processing the contents. 

\begin{figure*}[t]
    \centering
    \includegraphics[width=\textwidth]{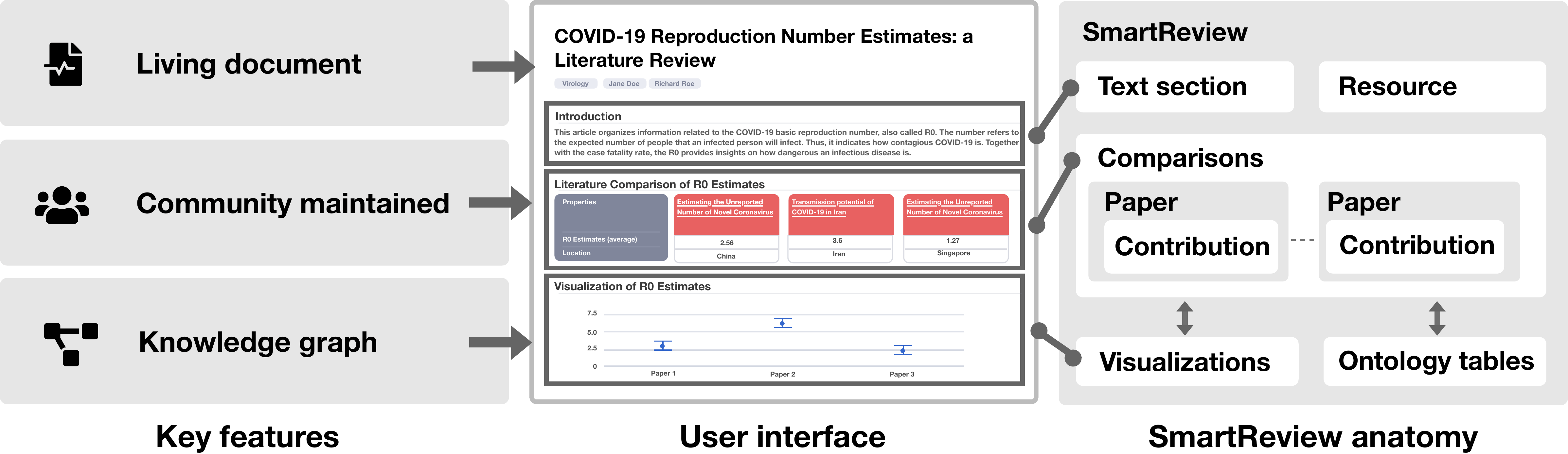}
    \caption{Illustration of key features and anatomy of SmartReviews. They are composed of several building blocks, including natural text, comparisons and visualizations.}
    \label{figure:teaser}
\end{figure*}

In this work, we present SmartReviews, a novel tool to author and publish review articles. The tool implements the requirements from the equally named SmartReview approach~\cite{oelenSmart2021Fixed} which addresses the weaknesses from which current review articles are suffering. Reviews are authored in a community-based manner and are represented as living documents, meaning that they can be updated whenever deemed necessary by the community. SmartReviews are implemented within an existing scholarly knowledge graph called Open Research Knowledge Graph (ORKG)~\cite{jaradeh2019open}. The key features and anatomy of SmartReviews are depicted in Fig.~\ref{figure:teaser}. In summary, this article provides the following research contributions:
\begin{enumerate*}[label=(\roman*)]
\item Detailed description of authoring and publishing semantic review articles using knowledge graphs.
\item Implementation of SmartReview authoring tool. 
\item Presentation and evaluation of an original SmartReview article.
\end{enumerate*}

\section{Related Work}
\label{section:related-work}
The current review authoring and publishing method faces numerous limitations and weaknesses~\cite{Oelen2020a}. In recent work~\cite{oelenSmart2021Fixed}, we identified these limitations and described them in detail. Table~\ref{table:weaknesses} summarizes them and includes an extended list of supporting related work. Based on those weaknesses, we devised an approach to address them. The two most pressing weaknesses relate to the inability to update articles once published and to the machine-inactionability of the presented knowledge. Both of these topics are extensively discussed in the literature. 

Shanahan advocates for ``living documents'' and to move away from the traditional and obsolete print model in which articles are sealed after publishing~\cite{Shanahan2015}. The living documents concept also provides opportunities for article retractions and corrections~\cite{Barbour2017}. This gives the possibility to embrace the features the modern web has to offer, including semantic web technologies~\cite{Shotton2009a}. Berners-Lee et al. used to term Linked Data to describe the interlinking of resources (i.e., data) by means of global identifiers, which constitutes the semantic web~\cite{berners2001semantic}. The Resource Description Framework (RDF) is the language used to represent the resources and provides an actionable format for machines~\cite{manola2004rdf}. RDF can be queried using the SPARQL query language~\cite{prudhommeaux2008sparql}. The use of these technologies improves the machine-actionability of data and provides a means to make data FAIR (Findable, Accessible, Interoperable, Reusable)~\cite{Wilkinson2016}. Semantic web technologies also play a key role in the living documents concept presented by Garcia-Castro et al.~\cite{Garcia-Castro2010}. This type of document supports tagging and interlinking of individual article components and embeds ontologies in the core of their approach.

\setlength{\tabcolsep}{4pt}
\begin{table}[t]
\centering
\caption{Summarized weaknesses of the current review and their respective related work. A detailed list of the weaknesses is presented in previous work~\cite{oelenSmart2021Fixed}.}
\label{table:weaknesses}
\begin{adjustbox}{scale=.9}
\begin{tabular}{p{3cm}|p{7.5cm}|p{2.2cm}}
\toprule
\textbf{Weakness} & \textbf{Definition} & \textbf{Related work} \\ \midrule
Lacking updates & Published articles are generally not updated due to technical limitations or lacking author incentives & \cite{Oelen2020a,MENDES2020110607,Meyer2011} \\ \hline
Lacking collaboration & Only the viewpoint from the review authors is reflected and not from the community as a whole & \cite{Oelen2020a,schmidt2005mites} \\ \hline
Limited coverage & Reviews are only conducted for popular fields and are lacking for less popular ones & \cite{Oelen2020a,Wee2016,Webster2002} \\ \hline
Lacking machine-actionability & The most frequently used publishing format is PDF, which hinders machine-actionability & \cite{Oelen2020a,Klampfl2014,Nasar2018,Lipinski2013,Jung2017,Heidorn2008a,Correa2017a} \\ \hline
Limited accessibility & The articles in PDF format are often inaccessible for readers with disabilities & \cite{Ahmetovic2018,Darvishy2018Fixed,NGANJI2015254} \\ \hline
Lacking overarching systematic representation & Web technologies are not used to their full potential because systematic representations are often lacking & \cite{Oelen2020a,Shotton2009a} \\ \bottomrule
\end{tabular}
\end{adjustbox}
\end{table}

\section{Approach}
\label{section:approach}

Our approach addresses the previously listed weaknesses. Accordingly, we introduce dimensions to address each weakness individually. The dimensions comprise: \begin{enumerate*}[label=(\roman*)]
\item Article updates
\item Collaboration
\item Coverage
\item Machine-actionability 
\item Accessibility 
\item Systematic representation
\end{enumerate*}. The approach leverages the SmartReview requirements as presented in~\cite{oelenSmart2021Fixed}.

The ORKG is used at the core of our approach. The use of knowledge graphs enables the reuse of existing ontologies, thus improving the machine-actionability of the data. To this end, the article has to be represented in a structured and semantic manner. Research articles are generally composed of multiple (non-structured) artifacts, among others this includes natural text sections, figures, tables, and equations. Review articles, in particular, do often include an additional artifact in the form of comparison tables. These tables present the reviewed work in a structured manner and compare the work based on a set of predefined properties. A previous study indicated that approximately one out of five review articles contains such tables~\cite{10.1007/978-3-030-64452-9_35}. Due to the structured nature of comparison tables, they can be processed more easily by machines. Complemented with semantic descriptions of the data, the comparisons can become FAIR data~\cite{Oelen2020a}. Therefore, we use comparison tables as the basis of our SmartReview approach. We leverage the comparisons tables within the ORKG which are specifically designed to be machine-actionable. 
\section{Implementation}
\label{section:implementation}

\begin{figure}
    \centering
    \includegraphics[width=1\textwidth]{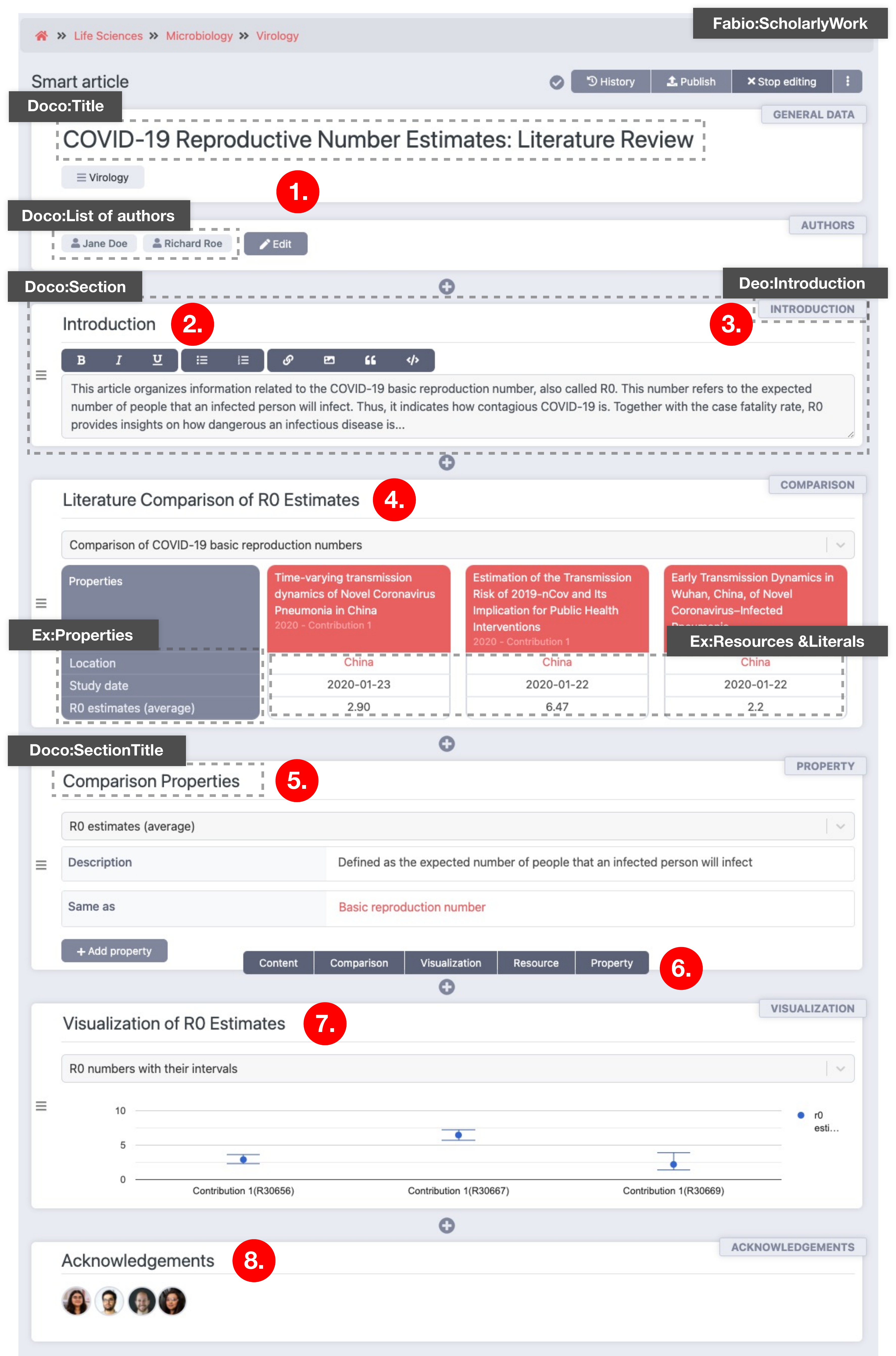}
    \caption{Screenshot of the implemented interface. Black labels represent the RDF types. Types prefixed with ``Ex'' are from the scholarly graph used for the implementation. Node 1 relates to the metadata of the article. Node 2 is the natural text content section and its Markdown editor. Node 3 shows the DEO type, which can be selected by the users when clicking on the label. Node 4 is a comparison section and node 5 is a property section. Node 6 shows the type selection for a new section. Node 7 is the visualization of the comparison shows in node 4. Finally, node 8 lists the contributors of the article.
    }
    \label{figure:interface}
\end{figure}

The interface is implemented in JavaScript using the React framework, the source code is available online\footnote{\url{https://gitlab.com/TIBHannover/orkg/orkg-frontend/-/tree/master/src/components/SmartReview}}. Additionally, a feature demonstration video is available.\footnote{\url{https://doi.org/10.5446/53601}} The knowledge graph is built on top of a Neo4j property graph and SPARQL support is provided via a Virtuoso endpoint.

\subsection{Section Types}

The main building blocks of SmartReviews are sections. Each section has a section type that describes the section's content and its relation to the knowledge graph. The article writer has been implemented on top of the ORKG which allows for reusing artifacts already present in the graph. When adding a section, the type can be selected (Fig.~\ref{figure:interface}, node 6). The section types comprise:
\begin{itemize}
    \item \textbf{Natural language text} sections support markup text via a Markdown editor. References are supported via a custom syntax using the same notation as R Markdown~\cite{xie2018r}.
    \item \textbf{Comparison} sections consist of tabular overviews of scientific contributions from papers that are being compared based on a selected set of properties. Comparison sections form the core of each review article.
    \item \textbf{Visualization} sections provide visual views of comparison data.
    \item \textbf{Ontology table} sections list descriptions of the properties and resources used within comparisons.
    \item \textbf{Resource and property} sections show a tabular representation of used resources and properties and their definitions from the knowledge graph.
\end{itemize}

\subsection{Implementation per SmartReview Dimension}
The implementation consists of various components to provide a comprehensive authoring interface. Among other things, this includes support for in-text citations, an interactive article outline, and reading time estimation. These features are ordinary functionalities for authoring tools and are therefore not discussed in detail. In the remainder of this section, we specifically focus on the dimensions of the SmartReview approach since they form the basis of the implementation. 

\textbf{Article updates.}
The requirement of updating articles combined with the requirement to keep persistent versions of articles introduces a level of versioning complexity. Especially due to the interlinking of knowledge graph resources, persistency is a complex endeavor that requires versioning at the resource level. To reduce the complexity, we added the constraint that only the latest version of an article can be updated, which we call the \textit{head version}. The head version is the only version that is stored in the graph. This implies that always the latest version of the article is present in the graph, leaving version complexity outside the graph and thus making it easier to interact with the graph. When an article is published within the system (not to be confused with publishing the article via a publisher), a snapshot is created of the subgraph used to generate the article. This approach resembles that of other collaborative curation systems (such as Wikipedia) that only allow edits of the latest version and keep a persistent history of all versions. Crucial for this approach is the ability to compare previous versions and to track individual changes (i.e., the diff view). 

\textbf{Collaboration.}
Collaboration is supported by allowing edits from any user. As with the article updates, this resembles the approach Wikipedia takes to support collaborative authoring. In Wikipedia, this has resulted in high-quality articles, which is popularly explained by the ``wisdom of the crowd'' principle~\cite{So2008}. To acknowledge researchers who contributed to the article, and to create an incentive to contribute, the acknowledgements section automatically lists anyone involved in writing the article (Fig.~\ref{figure:interface}, node 8). The list of acknowledgements is generated by traversing the article subgraph.

\textbf{Coverage.}
The only prerequisite to be able to contribute to an article is the need for a user account. Authentication serves for tracking provenance data (needed for the acknowledgements) and as a basic abuse prevention system.

\textbf{Machine-actionability.}
As described, the article content is available in the knowledge graph. The data itself can be accessed via various methods, including a SPARQL endpoint, RDF dump, and REST interface. To enhance machine interoperability, (scholarly) publishing ontologies were used. In Fig.~\ref{figure:interface}, RDF types prefixed with their respective ontologies are displayed next to system components. This includes the Document Components Ontology (DOCO)\footnote{\label{footnote:spar} \url{http://purl.org/spar/{doco,fabio,deo}}} to describe documents components. The FRBR-aligned Bibliographic Ontology (Fabio)\footnotemark[\getrefnumber{footnote:spar}] to describe the types of published work and the Discourse Elements Ontology (DEO)\footnotemark[\getrefnumber{footnote:spar}] ontology for specifying the section types. For the latter ontology, the article authors are responsible to select the appropriate type from a list of all DEO types for natural text sections (Fig.~\ref{figure:interface}, node 3). 

\textbf{Accessibility.}
Review articles are available as HTML files, which makes them by design more accessible than their PDF counterpart. Furthermore, WCAG guidelines are followed to enhance accessibility. In particular, semantic HTML tags are used as well as hierarchical headings. Finally, articles are responsive (i.e., support different screen sizes) making them suitable for high browser zoom levels and large font settings. 

\textbf{Systematic representation.}
Comparison tables form the main component to support systematic representations (Fig.~\ref{figure:interface}, node 4). The tables are created in a spreadsheet-like editor. The papers used within the comparison are represented as structured data in the graph, including the metadata such as title, authors, and publication date. Furthermore, the properties and their corresponding values are stored in the graph. When creating the comparison table, users are recommended to use existing properties and resources to further enhance interlinking. 

\section{Evaluation}
\label{section:evaluation}
To evaluate our approach, we now present a use case with an original SmartReview article to demonstrate how SmartReviews look like and how they differ from regular reviews. Afterwards, we demonstrate how data presented within the article can be accessed in a machine-actionable manner.

The SmartReview presents a selective literature review, titled ``Scholarly Knowledge Graphs'', and it published online\footnote{\url{https://www.orkg.org/orkg/smart-review/R135360}}. It consists of three comparisons and reviews in total 14 articles related to various types of scholarly knowledge graphs (i.e., identifier, bibliographic, domain-specific systems). This use case highlights the differences with regular static review articles. While regular review articles generally review the literature in comprehensive (and possibly lengthy) text sections, the SmartReview example shows how, instead, comparison tables are used to compare literature. Due to the interactive nature of the tables, they can contain more information than tables presented in static PDF files. Another notable difference is the presence of ontology tables within the article. The benefit of such tables is twofold: They improve machine-readability by linking the used properties to existing ontologies and improve human comprehension by textually describing the meaning of the property. 

To demonstrate the machine-actionability of SmartReviews, we now present four SPARQL queries that are used to query the underlying data (cf. Query \ref{query:sparql1}, \ref{query:sparql2}, \ref{query:sparql3}, and \ref{query:sparql4}). The first query is for metadata, whereas the other queries are for the actual knowledge presented in the respective articles. The prefixes \textit{orkgc}, \textit{orkgp} and \textit{orkgr} represent the class, predicate and resource URIs respectively.

\hspace{-0.5cm}

\begin{minipage}[t]{0.48\textwidth}
\begin{adjustbox}{scale=.85}
\begin{lstlisting}[captionpos=b, caption=Return all SmartReviews with research field (P30) information science (R278)., label=query:sparql1,
   basicstyle=\ttfamily\footnotesize,frame=single,showlines=true]
SELECT DISTINCT ?smartReview
WHERE {
  ?smartReview a orkgc:SmartReview;
       orkgp:P30 orkgr:R278.
}

\end{lstlisting}
\end{adjustbox}
\end{minipage}
\hspace{0.15cm}
\begin{minipage}[t]{0.48\textwidth}
\begin{adjustbox}{scale=.85}
\begin{lstlisting}[captionpos=b, caption=Return paper contributions (P31) addressing Scholarly Communication (R49584) as research problem (P32)., label=query:sparql2,
   basicstyle=\ttfamily\footnotesize,frame=single,showlines=true]
SELECT DISTINCT ?paper
WHERE {
  ?contrib a orkgc:Contribution;
       orkgp:P32 orkgr:R49584.
  ?paper orkgp:P31 ?contrib.
}
\end{lstlisting}
\end{adjustbox}
\end{minipage}

\begin{minipage}[t]{0.48\textwidth}
\begin{adjustbox}{scale=.85}
\begin{lstlisting}[captionpos=b, caption={Return all introduction sections from SmartReviews related to information science (R278).}, label=query:sparql3,
   basicstyle=\ttfamily\footnotesize,frame=single]
SELECT DISTINCT ?section
WHERE {
  ?review a orkgc:SmartReview;
       orkgp:P27 orkgr:R8193;
       orkgp:P31 ?contrib.
  ?contrib orkgp:HasSection ?section.
  ?section a orkgc:Introduction.
}
\end{lstlisting}
\end{adjustbox}
\end{minipage}
\hspace{0.15cm}
\begin{minipage}[t]{0.48\textwidth}
\begin{adjustbox}{scale=.85}
\begin{lstlisting}[captionpos=b, caption=Return all scholarly communication systems (R49584) with RDF support (P7009)., label=query:sparql4,
   basicstyle=\ttfamily\footnotesize,frame=single,showlines=true]
SELECT DISTINCT ?paper
WHERE {
  ?contrib a orkgc:Contribution;
       orkgp:P32 orkgr:R49584;
       orkgp:P7009 "T"^^xsd:string.
  ?paper orkgp:P31 ?contrib.
}

\end{lstlisting}
\end{adjustbox}
\end{minipage}

\section{Discussion}
\label{section:discussion}
We acknowledge that our proposed approach is radical and will unlikely be immediately adopted in every aspect by the research community. While some of the weaknesses originate from technology limitations, the main challenge is not technological in nature. Rather it is rooted in researchers' habits and mindsets and being comfortable with familiar methods. This relates to the open access movement~\cite{Bateman2006} which does not face a technical challenge but complex change that involves many aspects of traditional publishing. 

Our proposed approach does not solely address review authoring but also impacts the publication and dissemination process. Articles can be published and accessed via the platform's user interface or directly via the graph. Therefore, the platform serves as a digital library for review articles. As discussed, any user can author new articles and contribute to existing articles. This means that articles are not peer-reviewed in the traditional sense, rather a community-based continuous review method is performed. However, traditional peer-review is still possible. For example, as soon as an article is mature enough (which is decided by the authors), it can be published with traditional publishing means. However, we want to stress that a traditional publishing body is optional and is therefore not part of our approach. 

An extensive user evaluation is required to access the interactions and actual use of the system. Additionally, this user evaluation can focus on the usability aspects of the system. For future work, we have planned an evaluation with domain experts who will be asked to create a SmartReview for their field of expertise. This includes the creation of relevant comparisons and visualizations. 

Our approach can be generalized to research articles. Concretely it means that the article writer can be used to author any type of scholarly article. We focused on review articles because several of the weaknesses are most apparent for this type of article. Furthermore, we deem the limitation of static non-updated articles as a key limitation for reviews. 

\section{Conclusions}
\label{section:conslusion}
We presented the SmartReview tool, an application to author and publish scholarly review articles in a semantic and community-maintained manner. With the implementation, we address the current weaknesses of review article authoring and demonstrate a possible future of publishing review articles. A scholarly knowledge graph is used at the core of our approach, which increases the machine-actionability of the presented knowledge. The presented use case demonstrates how SmartReviews look like and it shows that the contents within articles is published in a machine-actionable manner. 

\subsubsection*{Acknowledgements}
This work was co-funded by the European Research Council for the project ScienceGRAPH (Grant agreement ID: 819536) and the TIB Leibniz Information Centre for Science and Technology.

\newpage

\bibliographystyle{splncs04}
\bibliography{references-mendeley,manual-references} %

\begin{thebibliography}{10}
\providecommand{\url}[1]{\texttt{#1}}
\providecommand{\urlprefix}{URL }
\providecommand{\doi}[1]{https://doi.org/#1}

\bibitem{Ahmetovic2018}
Ahmetovic, D., Armano, T., Bernareggi, C., Berra, M., Capietto, A., Coriasco,
  S., Murru, N., Ruighi, A., Taranto, E.: {Axessibility: A LaTeX package for
  mathematical formulae accessibility in PDF documents}. ASSETS 2018 -
  Proceedings of the 20th International ACM SIGACCESS Conference on Computers
  and Accessibility pp. 352--354 (2018). \doi{10.1145/3234695.3241029}

\bibitem{Barbour2017}
Barbour, V., Bloom, T., Lin, J., Moylan, E.: {Amending published articles: time
  to rethink retractions and corrections?} F1000Research  \textbf{6}, ~1960
  (2017). \doi{10.12688/f1000research.13060.1}

\bibitem{berners2001semantic}
Berners-Lee, T., Hendler, J., Lassila, O.: {The semantic web}. Scientific
  american  \textbf{284}(5),  34--43 (2001)

\bibitem{Correa2017a}
Corr{\^{e}}a, A.S., Zander, P.O.: {Unleashing Tabular Content to Open Data} pp.
  54--63 (2017). \doi{10.1145/3085228.3085278}

\bibitem{Darvishy2018Fixed}
Darvishy, A.: {PDF accessibility: Tools and challenges}, vol. 10896 LNCS.
  Springer International Publishing (2018). \doi{10.1007/978-3-319-94277-3\_20}

\bibitem{Garcia-Castro2010}
Garcia-Castro, A., Labarga, A., Garcia, L., Giraldo, O., Monta{\~{n}}a, C.,
  Bateman, J.A.: {Semantic Web and Social Web heading towards Living Documents
  in the Life Sciences}. Journal of Web Semantics  \textbf{8}(2-3),  155--162
  (2010). \doi{10.1016/j.websem.2010.03.006}

\bibitem{Heidorn2008a}
Heidorn, P.B.: {Shedding light on the dark data in the long tail of science}.
  Library Trends  \textbf{57}(2),  280--299 (2008)

\bibitem{jaradeh2019open}
Jaradeh, M.Y., Oelen, A., Farfar, K.E., Prinz, M., D'Souza, J., Kismih{\'{o}}k,
  G., Stocker, M., Auer, S.: {Open research knowledge graph: next generation
  infrastructure for semantic scholarly knowledge}. In: Proceedings of the 10th
  International Conference on Knowledge Capture. pp. 243--246 (2019).
  \doi{10.1145/3360901.3364435}

\bibitem{Jinha2010a}
Jinha, A.: {Article 50 million: An estimate of the number of scholarly articles
  in existence}. Learned Publishing  \textbf{23}(3),  258--263 (2010).
  \doi{10.1087/20100308}

\bibitem{Jung2017}
Jung, D., Kim, W., Song, H., Hwang, J.I., Lee, B., Kim, B., Seo, J.:
  {ChartSense: Interactive data extraction from chart images}. Conference on
  Human Factors in Computing Systems - Proceedings  \textbf{2017-May},
  6706--6717 (2017). \doi{10.1145/3025453.3025957}

\bibitem{So2008}
Kittur, A., Kraut, R.E.: {Harnessing the Wisdom of Crowds in Wikipedia: Quality
  Through Coordination}  (2008)

\bibitem{Klampfl2014}
Klampfl, S., Granitzer, M., Jack, K., Kern, R.: {Unsupervised document
  structure analysis of digital scientific articles}. International Journal on
  Digital Libraries  \textbf{14}(3-4),  83--99 (2014).
  \doi{10.1007/s00799-014-0115-1}

\bibitem{Lipinski2013}
Lipinski, M., Yao, K., Breitinger, C., Beel, J., Gipp, B.: {Evaluation of
  header metadata extraction approaches and tools for scientific PDF
  documents}. Proceedings of the ACM/IEEE Joint Conference on Digital Libraries
  pp. 385--386 (2013). \doi{10.1145/2467696.2467753}

\bibitem{manola2004rdf}
Manola, F., Miller, E., McBride, B., {others}: {RDF primer}. W3C recommendation
   \textbf{10}(1-107), ~6 (2004)

\bibitem{MENDES2020110607}
Mendes, E., Wohlin, C., Felizardo, K., Kalinowski, M.: {When to update
  systematic literature reviews in software engineering}. Journal of Systems
  and Software  \textbf{167},  110607 (2020). \doi{10.1016/j.jss.2020.110607}

\bibitem{Meyer2011}
Meyer, C.A.: {Distinguishing published scholarly content with CrossMark}.
  Learned Publishing  \textbf{24}(2),  87--93 (2011). \doi{10.1087/20110202}

\bibitem{Nasar2018}
Nasar, Z., Jaffry, S.W., Malik, M.K.: {Information extraction from scientific
  articles: a survey}, vol.~117. Springer International Publishing (2018).
  \doi{10.1007/s11192-018-2921-5}

\bibitem{NGANJI2015254}
Nganji, J.T.: {The Portable Document Format (PDF) accessibility practice of
  four journal publishers}. Library {\&} Information Science Research
  \textbf{37}(3),  254--262 (2015).
  \doi{https://doi.org/10.1016/j.lisr.2015.02.002}

\bibitem{Oelen2020a}
Oelen, A., Jaradeh, M.Y., Stocker, M., Auer, S.: {Generate FAIR Literature
  Surveys with Scholarly Knowledge Graphs}. JCDL '20: Proceedings of the
  ACM/IEEE Joint Conference on Digital Libraries in 2020 pp. 97--106 (2020).
  \doi{10.1145/3383583.3398520}

\bibitem{10.1007/978-3-030-64452-9_35}
Oelen, A., Stocker, M., Auer, S.: {Creating a Scholarly Knowledge Graph from
  Survey Article Tables}. In: Ishita, E., Pang, N.L.S., Zhou, L. (eds.) Digital
  Libraries at Times of Massive Societal Transition. pp. 373--389. Springer
  International Publishing, Cham (2020)

\bibitem{oelenSmart2021Fixed}
Oelen, A., Stocker, M., Auer, S.: {SmartReviews: Towards Human- and
  Machine-Actionable Reviews}. In: Linking Theory and Practice of Digital
  Libraries, Proceedings of TPDL 2021, pp. 181--186 (2021).
  \doi{10.1007/978-3-030-86324-1\_22}

\bibitem{prudhommeaux2008sparql}
Prudhommeaux, E., Seaborne, A.: {SPARQL query language for RDF}  (2008),
  \url{http://www.w3.org/TR/rdf-sparql-query/}

\bibitem{Randolph2009a}
Randolph, J.J.: {A guide to writing the dissertation literature review}.
  Practical Assessment, Research and Evaluation  \textbf{14}(13) (2009)

\bibitem{schmidt2005mites}
Schmidt, L.M., Gotzsche, P.C.: {Of mites and men: reference bias in narrative
  review articles; a systematic review}. Journal of family practice
  \textbf{54}(4),  334--339 (2005)

\bibitem{Shanahan2015}
Shanahan, D.R.: {A living document: reincarnating the research article}. Trials
   \textbf{16}(1), ~151 (2015). \doi{10.1186/s13063-015-0666-5},
  \url{https://doi.org/10.1186/s13063-015-0666-5}

\bibitem{Shotton2009a}
Shotton, D.: {Semantic publishing: The coming revolution in Scientific journal
  publishing}. Learned Publishing  \textbf{22}(2),  85--94 (2009).
  \doi{10.1087/2009202}

\bibitem{Webster2002}
Webster, J., Watson, R.T.: {Analyzing the Past to Prepare for the Future:
  Writing a Literature Review.} MIS Quarterly  \textbf{26}(2),  xiii -- xxiii
  (2002). \doi{10.1.1.104.6570}

\bibitem{Wee2016}
Wee, B.V., Banister, D.: {How to Write a Literature Review Paper?} Transport
  Reviews  \textbf{36}(2),  278--288 (2016).
  \doi{10.1080/01441647.2015.1065456}

\bibitem{Wilkinson2016}
Wilkinson, M.D., Dumontier, M., Aalbersberg, I.J., Appleton, G., Axton, M.,
  Baak, A., Blomberg, N., Boiten, J.W., da~Silva~Santos, L.B., Bourne, P.E.,
  Bouwman, J., Brookes, A.J., Clark, T., Crosas, M., Dillo, I., Dumon, O.,
  Edmunds, S., Evelo, C.T., Finkers, R., Gonzalez-Beltran, A., Gray, A.J.,
  Groth, P., Goble, C., Grethe, J.S., Heringa, J., t~Hoen, P.A., Hooft, R.,
  Kuhn, T., Kok, R., Kok, J., Lusher, S.J., Martone, M.E., Mons, A., Packer,
  A.L., Persson, B., Rocca-Serra, P., Roos, M., van Schaik, R., Sansone, S.A.,
  Schultes, E., Sengstag, T., Slater, T., Strawn, G., Swertz, M.A., Thompson,
  M., Van Der~Lei, J., Van~Mulligen, E., Velterop, J., Waagmeester, A.,
  Wittenburg, P., Wolstencroft, K., Zhao, J., Mons, B.: {Comment: The FAIR
  Guiding Principles for scientific data management and stewardship}.
  Scientific Data  \textbf{3}, ~1--9 (2016). \doi{10.1038/sdata.2016.18}

\bibitem{Bateman2006}
Willinsky, J.: {The Access Principle: the case for open access to research and
  scholarship}. Lhs  \textbf{2}(1),  165--168 (2006)

\bibitem{Wolmark2001}
Wolmark, Y.: {Quality assessment}. Gerontologie et Societe  \textbf{99}(4),
  131--146 (2001). \doi{10.3917/gs.099.0131}

\bibitem{xie2018r}
Xie, Y., Allaire, J.J., Grolemund, G.: {R markdown: The definitive guide}. CRC
  Press (2018)

\end{thebibliography}

\end{document}